# Spiropyran Sulfonates for Photo and pH Responsive Air-Water Interfaces and Aqueous Foam

*Marco Schnurbus,*[1,2] *Malgorzata Kabat,*[3] *Ewelina Jarek*[3], *Marcel Krzan*[3], *Piotr Warszynski,*[3] *and Björn Braunschweig*[1,2]*

[1] Institute of Physical Chemistry, Westfälische Wilhelms-Universität Münster, Corrensstraße 28/30, 48149 Münster, Germany

[2] Center for Soft Nanoscience, Westfälische Wilhelms-Universität Münster, Busso-Peus-Straße 10, 48149 Münster, Germany

[3] Jerzy Haber Institute of Catalysis and Surface Chemistry, Polish Academy of Sciences, ul. Niezapominajek 8, 30-239 Krakow, Poland

AUTHOR INFORMATION

**Corresponding Author** *e-mail: braunschweig@uni-muenster.de.

ORCID

Björn Braunschweig: 0000-0002-6539-1693

**ABSTRACT**

Responsive foams and interfaces, are interesting building blocks for active materials that respond and adapt to external stimuli. We have used the photochromic reaction of a spiropyran sulfonate surfactant to render interfacial, rising bubble as well as foaming properties active to light stimuli. In order to address the air-water interface on a molecular level, we have applied sum-frequency generation (SFG) spectroscopy which has provided qualitative information on the surface excess and the interfacial charging state as a function of light irradiation and solution pH. Under blue light irradiation, the surfactant forms a closed ring spiro form (SP), whereas under dark conditions the ring opens and the merocyanine (MC) form is generated. Using SFG spectroscopy, we show that at the interface, different pH conditions of the bulk solution lead to changes in the interfacial charging state. We have exploited the fact that the MC surfactant's O-H group can be deprotonated as a function of pH, and used that to tune the molecules net charge at the interface. In fact, SFG spectroscopy shows that with increasing pH the intensity of the O-H stretching band from interfacial water molecules increases which we associate to an increase in surface net charge. At a pH of 5.3, irradiation with blue light leads to a reversible decrease of O-H intensities, whereas the C-H intensities were unchanged compared to the corresponding intensities under dark conditions. These results are indicative of changes in the surface net charge with light irradiation, which are also expected to influence the foam stability via changes in the electrostatic disjoining pressure. In fact, measurements of the foam stabilities are consistent with this hypothesis with higher foam stability under dark conditions. At pH 2.7 this behavior is reversed as far as the surface tension and surface charging as well as the foam stability are concerned. This is corroborated by rising bubble experiments, which demonstrated an unprecedented reduction of ~30 % in bubble velocity when the bubbles were irradiated with blue light compared to the velocity of bubbles with the surfactants in the dark state. Clearly, the light-triggered changes can be used to control foams, rising bubbles and fluid interfaces on a molecular level which renders them active to light stimuli.

**INTRODUCTION**

Responsive surfaces[1–5] and soft materials[6–11] that can change their properties as a function of an external stimulus such as light or temperature are interesting to develop new active and even adaptive materials. Such materials are capable of changing their properties like optical appearance, stability or their structure on demand to new situations if a feedback mechanism can be established. However, a prerequisite for the latter are highly responsive materials[12–19] and building blocks[20–22] such as fluid interfaces[23] that react to an external stimulus by massive changes in their physicochemical properties.[23–29] At a fluid interface, this can be achieved by ad- and desorption of molecules[30] and by charging and discharging of the interface[1] which can largely change electrostatic interactions at the interface as well as by significant molecular structure changes.[31] Such electrostatics are often used to render colloidal systems like foam or nanoparticle dispersions more stable as the long-range electrostatic disjoining forces increase when the interfaces in a colloidal system become highly charged.[32,33] In this work, we are focusing on aqueous foams as these are inherently interface-controlled systems with an established hierarchy from the molecular building blocks that are adsorbed at the air-water interface to thin foam films, bubbles, ensembles of bubbles and finally to the macroscopically visible foam.[1,32] The latter inherits the properties of the air-water interface through structure-property relations.[34,35] Thus, aqueous foams are an ideal playground to test responsive interface-active moieties that are used to tailor also responsive foam[17,18,23,25,36] as a model system for soft matter materials. In addition, foams also find important technological applications that range from lightweight materials, wastewater treatment to recycling processes. Consequently, foams are also targeted in many technologies as final products or intermediates. Such products are continuously attempted to fulfill the requirements of high quality, efficiency, health safety, usage of biodegradable materials[9,25] or low production costs just to mention a few. However, foaming is not always desired. In some applications, foam is usually much undesirable, as it can severely hinder the filling of containers, reduce the efficiency of reactions, and can cause surface defects in coatings. Therefore, foams are of scientific and industrial interest and many studies are dedicated to the understanding of their formation, their properties, stability and structure. In our work, we have investigated spiropyran and merocyanine surfactants[21,37–45] which can undergo photochromic reactions and are applied as new photoswitchable and also pH-responsive building block that can be activated by light and pH stimuli. In fact, we present the first study that addresses these surfactants on a molecular level at the air-water interface and their use as a stabilizer for aqueous foam. Using a multi-technique approach with complementary methods such as vibrational sum-frequency generation (SFG), surface tensiometry, the analysis of rising bubbles and foam stability measurements we were able to deduce structure-property relations from the interfacial molecular structure and charging state to the stability of macroscopic foam.

**EXPERIMENTAL DETAILS**

*Synthesis and sample preparation*

The synthesis of (E)-4-(2-(2-hydroxystyryl)-3,3-dimethyl-3H-indol-1-ium-1-yl)butane-1-sulfonate was performed following synthesis protocol that have been previously reported, [37,38] and is schematically shown in Figure 1. For a more detail description we refer to the Supplementary Information (SI). The chemicals for the synthesis were used without further purification. Ultrapure water (18.2 MΩcm; total oxidizable carbon ≤ 3 ppb; obtained from a *Merck* Milli-Q Reference A+ purification system) was used to prepare aqueous surfactant solutions. All samples were sonicated for at least 30 minutes. The pH was adjusted after preparation of the solution with NaOH (98%, *Honeywell*) or HCl (37% w/w, *VWR chemicals*) and determined with a pH meter (InoLab pH 720, *WTW*). Alconox solutions (*Sigma Aldrich*) were used to pre-clean all necessary glassware, which was subsequently dried and soaked in concentrated sulfuric acid (98%, *Carl Roth*) with NOCHROMIX (*Godax Labs*) for at least 12 hours. The acid cleaned equipment was rinsed with copious amounts of ultrapure water. All measurements reported below were performed at 295 K room temperature.

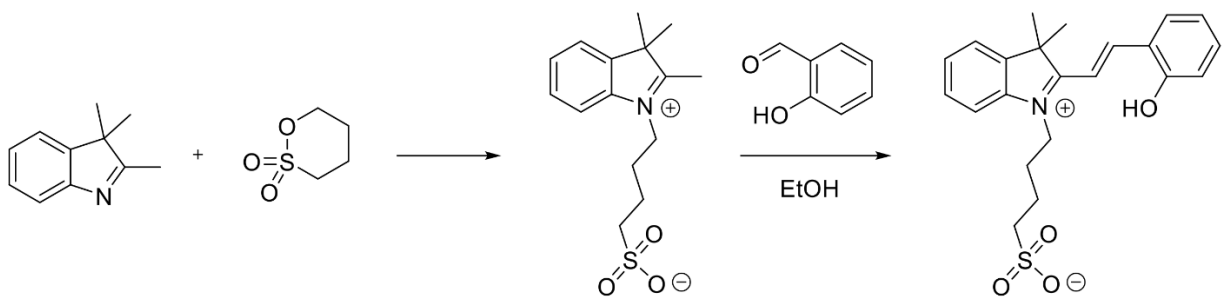

**Figure 1** *Synthesis route for the sulfonated spiropyran (SP).*

*Sum-Frequency Generation (SFG)*

Vibrational SFG spectroscopy is an inherently surface-specific method that is based on a second-order nonlinear effect. For SFG, two laser beams, a tunable broadband infrared and a frequency-fixed but narrowband visible beam are superimposed in time and space at the interface of interest, where they generate a third beam with the sum frequency (SF) of the two impinging beams. In general, the SF intensity is resonantly enhanced when the broadband IR beam excites molecular vibrations and can be expressed as follows[46,47]:

$$I_{SF} \propto \left| \chi_{NR}^{(2)} + \sum_q \frac{A_q}{\omega_q - \omega - i\Gamma_q} \right|^2 \qquad (1)$$

where $\chi_{NR}^{(2)}$, $\Gamma_q$, $\omega_q$ are a nonresonant contribution to the second-order susceptibility $\chi^{(2)}$, the Lorentzian linewidth, and the resonance frequency of the q$^{th}$ vibrational mode, respectively. In addition, $A_q \propto N\langle\beta_q\rangle$ is the SFG amplitude of the mode $q$, and is directly related to the number density of interfacial molecules and the orientational average of the molecules hyper-polarizability $\beta_q$. The orientational average $\langle\cdots\rangle$ of $\beta_q$ has far-reaching consequences as in dipole approximation it zeroes bulk signals of centrosymmetric materials such as liquids and gases in the time average. Besides, the above orientational average also provides a direct orientation analysis of a species by addressing (the phase or) the sign of the q$^{th}$ vibrational mode.[47–50] At interfaces of centrosymmetric materials, their bulk symmetric is necessarily broken which results into nonzero and purely interface related SFG signals. For that reason SFG spectroscopy can be a powerful tool to characterize the molecular composition, molecular order and charging state of interfaces. In addition, SFG signals in particular from interfacial water molecules (O-H stretching vibrations) at electrified interfaces has also a third-order contribution

$$I_{SF} \propto \left| \chi^{(2)} + \frac{\kappa}{\kappa + i\Delta\kappa_z} \chi_S^{(3)} \phi_0 \right|^2 \qquad (2)$$

that has, in fact, recently gained considerable interest and was frequently addressed in detail in previous works.[33,46,51] Here, this contribution depends on the double-layer potential $\phi_0$, the inverse Debye length $\kappa$ and the wave vector mismatch $\Delta\kappa_z$.

SFG experiments were performed with a setup that is described elsewhere.[33,52] Spectra were taken with the sum-frequency, the visible and the IR beams having s-, s- and p-polarization (ssp), respectively. The SFG spectra were recorded by tuning the center wavelength of the IR beam in 4 steps and were normalization to the nonresonant signal of an air-plasma cleaned polycrystalline Au surface.

*Surface Tensiometry – Drop Shape Analysis*

The surface tension $\gamma$ of surfactant modified air-water interfaces was investigated by applying the pendant drop method. In this experiment, the surface tension is determined by an analysis of the drop shape based on the application of the Young-Laplace equation.[53] This equation sets the difference in pressure (Laplace pressure) between the inside and outside of the drop in relation with the surface tension and the radii of curvature. For these experiments we have applied a PAT-1M (*Sinterface*, Germany) and a DSA100 (*Krüss*, Germany) tensiometer that were equipped with a modified cuvette with two long pass filters (*Schott* OG590) with a cutoff wavelength at 590 nm to restrict the wavelengths for image analysis of the drop shape to the red portion of the light source. Thus, the light used to determine the shape of the droplet did not affect the surfactant. Inside the cuvette, one LED with a center wavelength of 465 nm (spectrum in the Supporting Information) was placed and was used for in situ photo-switching, while the dynamic time-dependent changes in surface tension were simultaneously recorded until equilibrium was reached. For the surfactants under investigation, typically >30 min were needed to reach an equilibrium state at the air-water interface. For the measurements under dark conditions the LED was switched off and yet again the surface tension was recorded as a function of time until equilibrium was reached.

*Analysis of Rising Bubbles*

When a bubble rises in the surface-active agent solutions, an uneven distribution of surfactant molecules over its surface is induced as a result of a viscous drag exerted by the fluid on the moving bubble interface. Adsorption coverage is lowered on the upstream part, while an accumulation of adsorbed molecules of surface-active substance takes place at the rear part of the bubble. This gradient of the surface concentration reduces interfacial mobility and consequently, diminishes the velocity of the bubble. As a result of the bubble interface immobilization, the hydrodynamic drag for the bubble motion is increased, and the bubble rising velocity can be reduced, even by more than 50 % in the comparison with bubbles rising in pure liquids.[54] For complete immobilization of interface of the rising bubbles, a definite degree of adsorption coverage is needed, which magnitude depends on the chemical nature of the surface-active substance under investigation.[55–57] In previous works, it was shown that the local bubble velocities decrease with the increasing concentration in solution of surface-active, water-soluble species and rising bubble experiments thus interrogate the temporary state of a Dynamic Adsorption Layer DAL[55] and the hydrodynamic conditions in the bubble interface proximity.

Previously, it has been shown that the surface coverage and the dynamic structure of the adsorption layer, i.e., non-uniform distribution of the adsorbed molecules, are factors of crucial importance for the motion and thus the velocity of rising bubbles. The initial acceleration and the profiles of the local velocity of bubbles after detaching from the capillary orifice are strongly dependent on the degree of the initial surface coverage. At low surfactant concentrations, three stages of bubble motion - acceleration, deceleration and finally a steady-state (terminal velocity) – can be observed. At high concentrations, only the first and last stage occur and, thus, a local maximum of the bubble velocity is not observed. In general, a certain minimal surfactant coverage is needed for the disappearance of the maximum. This concentration decreases with increasing surface activity of the amphiphile.[58–60] For a precise determination of the local velocities and the acceleration of the bubbles which detached from the capillary, we have applied a modified setup that was already described previously.[61] In this applied setup, bubbles were generated in a 10 cm high glass column with a 30 mm x 30 mm cross-section that was filled with 30 ml of the solution. Bubbles were formed at controlled time intervals using a high precision peristaltic pump (ISMATEC ISM597D) that was connected to a glass capillary with an inner diameter of 75 μm. The size of the created bubbles was $1.48 \pm 0.01$ mm diameter in ultrapure water. The bubble formation time at the capillary orifice was $1.6 \pm 0.2$ s. The solution was equilibrated by irradiating with blue light that had a wavelength of 465 nm conditions or was kept in darkness for at least 10 min before each experiment. During this equilibration procedure, the sample solution was continuously mixed. The mixing was stopped and the solution was allowed to rest just before the rising bubble experiment was started. A digital ultra-high-speed camera (IDT Model NX3) recorded the bubble motion with frequency 4450 fps. The recording was performed under red light irradiation (595 nm high power diode). Note that the use of red light was necessary to avoid any unwanted light-induced changes in the surfactant isomeric form. The bubble local velocities were then automatically evaluated using available macros of ImageJ. The local bubble velocities $U$ were calculated using the coordinates of the subsequent positions of the bubble and the frequency of camera acquisition.

*Analysis of Macroscopic Foams*

In order to investigate the foaming properties of the surfactants in the merocyanine and spiropyran form, a dynamic foam analyzer (DFA100, *Krüss*, Germany) was used. This device allows to investigate the foamability as well as the foam stability and relies on an accurate measurement of the foam height as a function of time. Samples of the surfactants with a volume of 40 mL were loaded in a glass column (250 mm length and a diameter of 40 mm).

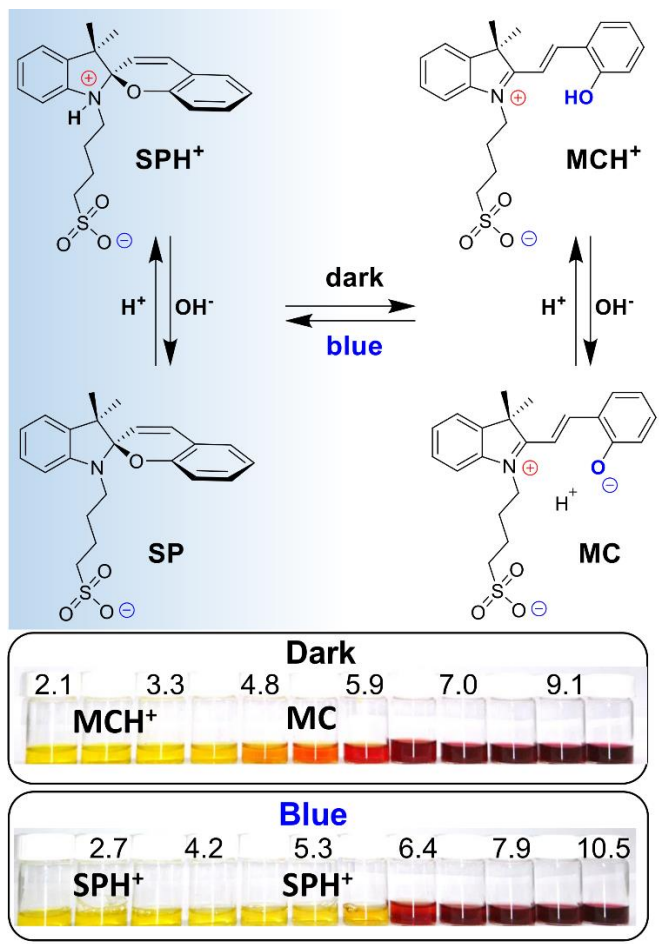

**Figure 2** *Structures of merocyanine (MC), protonated merocyanine (MCH⁺) and spiropyran (SP) as well as protonated spiropyran (SPH⁺) sulfonates as a function of pH and light irradiation. Counterions have been omitted for clarity. Note that at pH 5.3 the hydroxyl of the MC form is expected to be almost fully deprotonated (mixture of MC and MCH⁺), while for pH 2.7 only the MCH⁺ form is present.[21,38,62] In addition, photographs of the surfactant solutions for different pH values of the aqueous solution. The latter were either irradiated with 465 nm blue light or kept under dark conditions as indicated in the figure. pH values were as indicated in the photographs.*

To produce foams, ambient air was passed for 20 s with a flow rate of 0.5 L/min through a porous glass frit (P3, *Carl Roth*, Germany) which was fixed at the bottom of the column. The DFA100 was equipped with an IR LED panel (which does not affect the switching of the surfactant) at one side of the column and with a line photodiode detector at the side opposite to the LED panel. This setting allowed determining the foam height as a function of foam age by measuring the light transmission through the glass column. Solutions of spiropyran at concentrations of 4 mM (pH 2.7 and 5.3) were kept in an enclosed and light-tight chamber to equilibrate the samples in dark conditions for at least 30 minutes before the foaming experiment was started. The measurements with continuous light irradiation were done with 465 nm blue light and the solutions were first equilibrated for 20 minutes before the foaming experiment was initiated.

**RESULTS AND DISCUSSION**

The structure of spiropyran sulfonate surfactants is shown in Figure 2 as a function of pH and light conditions. The previous works[21,37,38,63] on similar molecules have pointed out that the thermodynamic more favorable merocyanine (MC) form is present at high pH, while the protonated merocyanine (MCH$^+$) forms at low pH, when the solutions are kept under dark conditions without light irradiation. This transition from the MCH$^+$ to the MC as a function of pH leads to a characteristic color change that is also seen in the photographs that are shown in Figure 2. From that we can attempt to estimate the pK$_a$ of hydroxyl of the MCH$^+$ to be between 5.9 and 6.4. After irradiation with blue light (465 nm), the molecule was switched via the photochromic reaction that is schematically shown in Figure 2 from the MC/MCH$^+$ conformation to the spiropyran (SP) form when the pH is high enough. However, at intermediate pH <6 and low pH the protonated spiropyran (SPH$^+$) is likely to be formed. The proposed reaction mechanism is also shown in Figure 2 for two different pH conditions, for which we will provide some experimental evidence below. For that, we have analyzed both the surface and the bulk properties at two different pH values, which were adjusted to 2.7 and 5.3. In the bulk solution, UV/Vis absorption spectra which are shown in Figure S1 of the Supporting Information were recorded and reveal a strong absorption band at 421 nm and two bands in the UV region at center wavelengths of 252 and 215 nm when the samples were in the MC/MCH$^+$ conformation (dark state). After irradiation with blue light the absorption band centered at 421 nm decreased, whereas the absorption of the two UV bands increased. Irradiation with blue light also leads to substantial changes in the color and, thus, to the appearance of the surfactant solutions, which can be directly seen by visual inspections as shown in Figure 2. These changes in optical properties are indicative for the photochromic reaction of the surfactants MC/MCH$^+$ (dark) to the SP/SPH$^+$ (blue light), which is highly reversible as we could perform many switching cycles without noticeable changes in the absorbance spectra (see Figure S1, Supporting Information).

In Figure 3, we present SFG spectra of the air-water interface for different surfactant concentrations that were taken after equilibration either in the dark or under constant blue light irradiation. Figure 3a presents SFG spectra where the bulk pH was fixed to 5.3, while in Figure 3b SFG spectra from the air-water interface at a solution pH of 2.7 are shown. For both pH values, broad vibrational bands at 3200 and 3450 cm$^{-1}$ due to O-H stretching vibrations of interfacial water molecules dominate the SFG spectra at high surfactant concentrations, while a much weaker vibrational band centered at 3057 cm$^{-1}$ can be attributed to the aromatic C-H stretching vibration. Additional bands at 2947, 2879 and 2856 cm$^{-1}$ are due to CH$_3$ antisymmetric stretching vibrations as well as due to symmetric CH$_3$ and CH$_2$ stretching vibrations, respectively.

With decreasing surfactant concentration, we observe a decrease in the SFG intensity of both O-H and C-H bands until the intensities of all C-H bands are negligible and a new narrow band centered at 3700 cm$^{-1}$ appears. This band is attributable to dangling O-H groups at the air-water interface that have no hydrogen bonds and which point into the gas phase. Clearly, the appearance of this band and the resemblance of the SFG spectrum with that of a neat air-water interface[64,65] is indicative of a negligible surface excess of the surfactant for concentrations <0.5 mM. A close comparison of the SFG spectra recorded under dark conditions with those during blue light irradiation in Figure 3a, reveals that the O-H intensity had decreased substantially when the samples were irradiated with blue light which triggers the photochromic reaction from the MC/MCH$^+$ to the SPH$^+$ conformation as proposed in Figure 2.

At this point we recall that the O-H intensities in SFG spectra can provide information on the interfacial charging state (see details section), which means that the differences in O-H intensity indicate a higher surface charge if the surfactants are kept in the dark at pH 5.3.[33,46,51] In addition, we also recall that the phase of a SFG active vibrational band is dominated by the net orientation of the interfacial species that gives rise to that band and therefore an analysis of the phase of the O-H stretching bands in SFG spectra can provide useful information on the net orientation of interfacial water molecules that orient themselves within the interfacial electric field.[46,48,51,66,67] From a visual inspection of all SFG spectra in Figure 3, we can conclude that the phase of the O-H stretching bands is not changing as the highly dispersive line shape of the aromatic C-H stretching band 3057 cm$^{-1}$ does not change as a function of concentration or pH. In fact, this band only varies in intensity together with the intensity of the broad O-H stretching modes. Note that the line shape of the 3057 cm$^{-1}$ band is consistent with an air-water interface having a negative net charge as a comparison with other systems shows. Such a comparison can be made with previous SFG studies on surfactant/polyelectrolyte[34] or surfactant dye mixtures,[35] where also aromatic residues were present, but the interfacial charging state was changed from a negative to a positively charged interface.

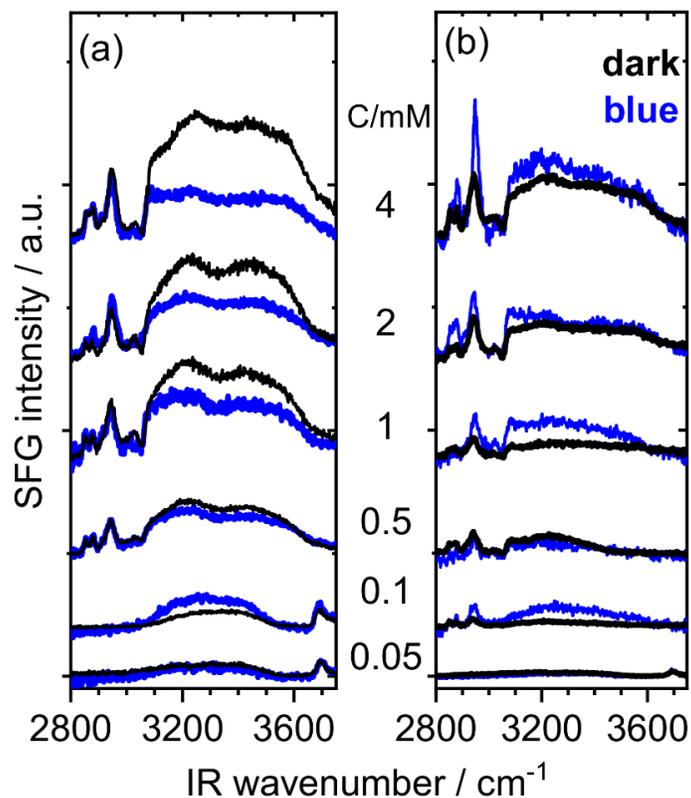

**Figure 3** *Vibrational SFG spectra of air-water interfaces modified with merocyanine (MC/MCH$^+$) sulfonate (black lines) and spiropyran (SPH$^+$) sulfonate (blue lines) surfactants which form under dark and blue light conditions, respectively. (a) Shows SFG spectra as a function of surfactant concentration when the solution pH was adjusted to 5.3, while (b) shows SFG spectra for solutions with the pH adjusted to 2.7. The surfactant concentrations were as indicated in the figure.*

From our analysis of the SFG spectra in Figure 3, we can conclude that the surface charge varies as a function of light irradiation, but the net orientation of interfacial water molecules is maintained at all concentrations >0.5 mM and for all pH values investigated.

We will now discuss the differences in interfacial properties for the two selected pH values (2.7 vs 5.3). Close inspection of the SFG spectra for 4 mM concentrations in Figures 3a and 3b, shows that the light-induced changes are different for the two pH values. In fact, for pH 5.3 and irradiation with blue light a decrease in O-H intensity and therefore in the interfacial charging state as discussed above is observed, but at pH 2.7 the behavior is opposite to what is seen at pH 5.3. In fact, for the latter pH we had observed a slight increase in the O-H intensity, when the samples were irradiated with blue light (Figure 3b).

As indicated in Figure 2, the MCH$^+$ form of the surfactant is at pH 2.7 and dark conditions zwitterionic with zero net charge. This leads to low SFG intensities from O-H stretching vibrations (Figure 3b). After irradiation with blue light, the surfactant switches from the MCH$^+$ to the protonated spiro (SPH$^+$) form, which is also zwitterionic and carries again no net charge (MCH$^+$ $\rightarrow$ SPH$^+$) and is consistent with similar O-H intensities and the absence of strong electric field-induced contributions for both pH values. The SPH$^+$ form at pH 2.7, however, seems to be more surface active as the MCH$^+$ form as evidenced by the strong rise in the intensity of vibrational C-H bands (Figure 3b), the sharp drop in surface tension (Figure 4b) as well as by the changes in local velocity of rising bubbles (Figure 4d). This can be rationalized by the higher hydrophilicity of the MCH$^+$ that is induced by the presence of the surfactants hydroxyl group which renders this spatial more demanding species (Figure 2) less surface active.

For surfactant solutions with pH of 5.3, we have observed pronounced changes in the O-H intensities. This can be explained in a similar line of arguments as in the case of pH 2.7, but at pH 5.3 the solution contains a mixture of MC and MCH$^+$ forms with a possible prevalence of the MC form.[21,38,62] This causes the MC form with a negative net charge (Figure 2) to dominate the air-water interface and results in a charged interface that is consistent with the much higher (electric field induced) O-H intensities in the SFG spectra of Figure 3a (pH 5.3, dark conditions). Irradiation with blue light now causes this higher negative net charge at the MC species to decrease when it is transferred to the zwitterionic SPH$^+$ form (Figure 2). This is accompanied by a substantial loss in SFG intensity from O-H stretching vibrations (MC → SPH$^+$).

In Figures 4a and 4b, we present the equilibrium surface tension for the two pH values of 5.3 and 2.7 as a function of concentration. Note that the highest concentration was just below the solubility limit. Clearly, the surface tension decreases with increasing concentration and was also changed with light irradiation. In fact, there is a noticeable but moderate decrease in surface tension for pH 5.3, while the decrease in surface tension after blue irradiation was for a pH of 2.7 substantial. In fact, at pH 5.3 the maximum change $\Delta\gamma$ in surface tension was ~8 mM/m, while for pH 2.7 the maximum change $\Delta\gamma$ in surface tension was ~18 mN/m.

Figures 4c and 4d present the profiles of the rising bubble local velocity at different light and pH conditions. In the Supporting Information we also present videos that were recorded for the rising bubbles. In these videos the different bubble velocities can be qualitatively inferred by a close comparison of rising bubbles with surfactants in the SPH$^+$ and MC form. The data in Figure 4c and 4d present the results of a detailed analysis of videos and show the initial bubble acceleration just after the detachment.

The bubble acceleration and the local velocity for the surfactant-laden solutions are lower at pH 2.7 as compared to pH 5.3 where little differences to the surfactant-free solutions (neat water) were found. From these observations, we can conclude that the surface-activity is much higher at pH 2.7, a fact that is also confirmed by surface tensiometry (Figures 4a and 4b) and SFG spectroscopy (see discussion above). For pH 2.7, massive changes occur when the samples were irradiated with blue light compared to dark conditions.

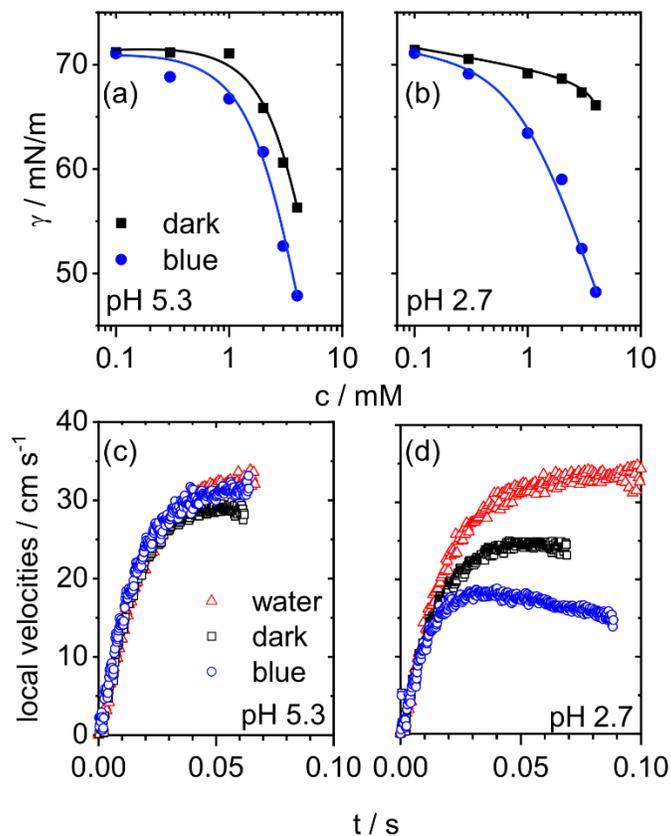

**Figure 4** *Equilibrium surface tension for merocyanine (MC) sulfonate (black squares) and spiropyran (SP) sulfonate (blue circles) surfactants which form under dark and blue light conditions, respectively and (a) pH 5.3 and (b) pH 2.7. (c) and (d) show the bubble velocity under different light and pH conditions and for 0.2 mM surfactant concentrations. Symbols for MC and SP forms are as in (a) and (b) in addition, we show the bubble velocity in neat water (red triangles). In the Supporting Information we show videos that highlight the different bubble velocities. Lines in (a) and (b) guide the eye.*

In fact, from a close inspection we can infer that the bubble velocity is dramatically reduced when the bubbles were irradiated by blue light and the surfactants in the SPH$^+$ conformation, while under dark conditions the bubble velocity is much lower than that for pure water but roughly 30 % higher as compared to the velocity of blue irradiated bubbles. This is a remarkable result and is unprecedented for a simple surfactant, where such dramatic changes in bubble velocity have been only reported for larger differences in surfactant concentrations e.g. by going from neat water to moderate surfactant concentrations.[54,55,59,68] We take these results as direct evidence that the surface of a rising bubble can be activated by a light stimulus when SP/MC surfactants are present. Due to the nature of the surfactants as discussed above the use SP/MC surfactants renders rising bubbles not only responsive to pH conditions which is caused by the protonation/deprotonation of the surfactants, but also to blue light irradiation that triggers conformational changes via photochromic reactions as shown in Figure 2. At this point, it is interesting to point out that the sensitivity of rising bubble experiment in terms of surfactant adsorption and variations of the surfactant surface activity is higher compared to classical tensiometry. [58–60] For pH 2.7 and blue light irradiation, we observed that the rising bubbles had a local plateau of the velocity at ~15 cm/s which is consistent with a fully immobilized surface shortly after the bubble was formed. This is directly connected with zwitterionic character of the surfactant that has a zero net charge and thus possesses a higher surface activity of the SPH$^+$ species at pH 2.7 and blue light irradiation. Clearly, a different behavior for pH 2.7 and dark condition where the MCH$^+$ conformation dominates is observed. In fact, in the latter case we can notice a higher maximum velocity (ca. 25 cm/s) and no deceleration in the reported distance of 20 mm from the capillary. This is indicative for a partial immobilization of the bubble interface, while in contrast, at pH 5.3 almost no effect of the presence of the surfactants is observed and the bubble rises with an acceleration similar to that of bubble in pure water. Clearly, for these conditions the excess of surface adsorbed molecules was not high enough to create a so-called dynamic adsorption layer[55] which can form immediately after bubble detachment. The observed effects of various redistributions of surface active substance in adsorption layer during light and pH variations are the similar as in the case of experiments with surfactants with different polar group and the same hydrocarbon chain. We have reported earlier that the minimum adsorption coverage of ~25 % is needed for a full immobilization of the bubble with n-octyltrimethylammonium bromide cationic surfactants, which was much larger compared to nonionic surfactants with $C_8$ alkyl chains.[59] That can be explained by the difference of surface activity between non-ionic surfactants and ionic ones induced by the effect of surface charge which accumulates during adsorption of ionic surfactants and produces an energy barrier for further adsorption.[69]

This analysis of rising bubbles, which are a necessary precursor for foams that are made up by ensembles of bubbles which first rise in a liquid column until they assemble above the liquid surface where they form the foam. The changes in bubble velocity that are associated with the changes in structure, coverage and charging state at the air-water interface can be used as a first estimate on the foam stability through structure-property relations from the air-water interface, to a rising bubble and finally to the macroscopic foam.

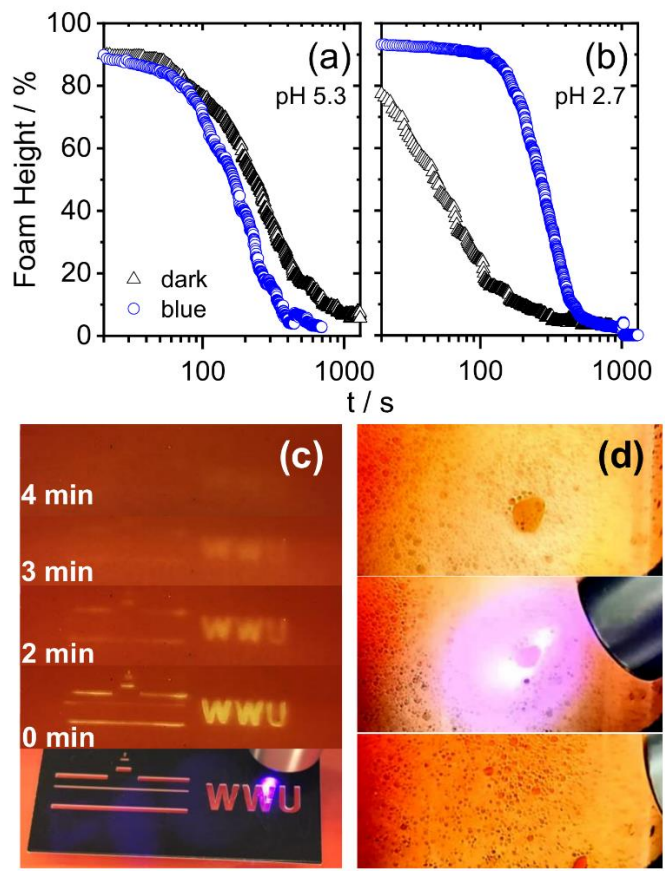

**Figure 5** *Comparison of the foam stability (height loss as a function of time) for aqueous foams from 4 mM spiropyran sulfonate solution with a bulk pH of (a) 5.3 and (b) 2.7. Blue circles indicate sample irradiation with blue light while black triangles show the foam height which was recorded under dark conditions. (c) Self-erasing information that is written into a 4 mM spiropyran sulfonate solution at pH 5.3 using a 405 nm laser (<1 mW), while (d) shows light on demand foam destruction in a MC/MCH$^+$ stabilized foam at a pH of 5.3. For (d) the same laser as in (c) was used. The videos of (c) and (d) are available online in the Supporting Information.*

For that we will now first report on our experimental results for aqueous foams from SPH$^+$ and MC/MCH$^+$ solutions. In Figure 5, we present the foam height as function of time for foams that were prepared from 4 mM surfactant solutions with a pH adjusted to 5.3 and 2.7. Foams from aqueous solutions with a pH of 5.3, showed a higher stability (longer foam lifetime), when the foam was kept in the dark (Figure 5a) while irradiation with blue light caused a slightly faster decay of the foam height and thus lower foam stabilities, and is possibly related to the interfacial charging state. This is clearly different from foams that were prepared from solutions with a pH of 2.7. Here, the foams were substantially more stable under blue light irradiation compared to the dark conditions (Figure 5b) and is likely caused by the high surface excess of SPH+ from and consequently the much lower surface tension at this pH and blue irradiation. Clearly, the trends in foam stability with pH and with light irradiation, qualitatively follow the changes in the interfacial charging state, surface tension, surface excess and rising bubble velocity which we have discussed above. We also point out that the concentration of 4 mM was the highest possible concentration due to solubility reasons and that the ionic strength was for pH 2.7 at 2.1 mM, whereas for pH 5.3 the ionic strength was only 0.2 mM for a 4 mM surfactant solution. These points may additionally account for but not dominate the differences in interfacial and foam properties from solution with pH 5.3 and 2.7.

Although, we can identify differences in foam stability as a function of the light irradiation e.g. at pH 2.7 the foam decay is much faster under dark conditions compared to the decay of foams that were irradiated with blue light, the relatively short lifetime of the SPH$^+$ or MC /MCH$^+$ stabilized foams, impairs their use as major responsive systems. For the latter ideally a very stable state is switched to an unstable state and vice versa. This has been shown for instance by Fameau et al.,[25,70] who have reported on foams from solutions that contained 12-hydroxystearic acid (12-HSA) molecules that were self-assembled into particles. The 12-HAS particles can facilitate ultra-stable foams, while temperature treatments can be used to destabilize these foams. This concept is similar to foams stabilized by thermo-responsive polymers such as hydroxypropyl cellulose.[9] Although, the lifetime of the spiropyrane and the merocyanine sulfonate stabilized foams presented in our study is short compared to the systems noted above, it is possible to change locally the color of the foam. This can be accomplished e.g. by going from the MC to the $SPH^+$ form, and results in a decrease of the optical absorbance for visible wavelengths as shown in Figure S1 of the Supporting Information where the solutions color appearance turns from dark red to light yellow (Figure 5c). In Figure 5c, we show how information can be imprinted in the solution, while this information is also self-erasing as the excited $SPH^+$ configuration returns to the thermodynamic more favorable $MC/MCH^+$ state after some time. Additionally, Figure 5d reports on an experiment where the foam at pH 5.3 was locally collapsed by irradiation using a low power (<1 mW) laser with a wavelength of 405 nm. Note that at this pH the foam was less stable for blue irradiation ($SPH^+$ form, Figure 4a). Although the differences between the $SPH^+$ and the $MC/MCH^+$ stabilized foams (blue vs dark conditions) were not as substantial as for pH 2.7 (Figure 5d), it is nonetheless possible to destroy the foam locally on demand.

**CONCLUSION**

We have investigated $MC/MCH^+$ and $SPH^+$ sulfonate surfactants at the air-water interface using vibrational sum-frequency generation spectroscopy and surface tensiometry and discussed their use as a new surfactant to render aqueous foams active to light stimuli via the photochromic reactions of the surfactants at the air-water interface and in the aqueous bulk solution. Using SFG spectroscopy we have addressed the changes in surface excess and surface charging state as a function of concentration, pH and light irradiation on a qualitative level and we have compared our conclusions to the stability of macroscopic foams. At the air-water interface, we have demonstrated that the surfactant can be switched between a merocyanine form, which stabilizes under dark conditions while the spiropyrane form is formed after irradiation with 465 nm blue light. To tune the net charge of the surfactant we changed the pH value: At a pH of 5.3 the O-H group of the $MCH^+$ form can be partially deprotonated and SFG spectroscopy showed that the intensity of O-H stretching bands from interfacial water molecules under dark conditions was increased substantially, which we associate to an increase in the surface net charge. At a pH of 2.7 we observed a different behavior where the O-H intensities were found to be slightly higher under blue light irradiation, while they decreased when the samples and interfaces were kept in the dark. The changes in foam stability showed an excellent correlation to the surface net charge and surface excess that we have gained from SFG spectroscopy and surface tensiometry. We propose to associate these changes with the variation of the disjoining pressure, which can be activated by light irradiation and can be tuned by pH adjustments. Clearly, structure-property relations from interfacial level to the macroscopic foam exist and can be used to tailor foam properties on a molecular level and to render foams responsive to light stimuli.

**SUPPORTING INFORMATION**

Supporting information on the synthesis procedures and the characterization of the synthesis products, UV/Vis spectroscopy of surfactant solutions including experimental details, videos of rising bubbles in solutions with different pH and under different light conditions.

**ACKNOWLEDGMENTS**

Financial support from a NAWA-DAAD bilateral program "Smart liquid/gas interfaces with photo-switchable surfactants" (Polish National Agency for Academic Exchange NAWA grant no. PPN/BIL/2018/1/00093, DAAD Projekt 57448918) and by the European Research Council (ERC) under the European Union's Horizon 2020 research and innovation program (grant agreement No. 638278), the National Science Center of Poland (grant no. 2016/21/B/ST8/02107) and the statutory funding of ICSC is gratefully acknowledged.

**Table of contents graphic (TOC)**

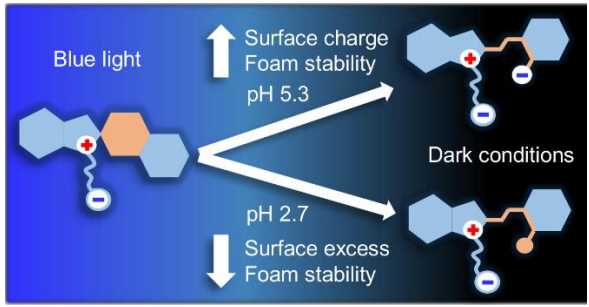